\documentclass{article}
\usepackage{amsmath, amsthm, amsfonts, amssymb, latexsym, graphicx}
\usepackage[margin=1in, letterpaper]{geometry}

\usepackage{tikz}
\usetikzlibrary{calc, through, arrows}

\newcommand{\mc}{\mathcal}

\newcommand{\p}{\prime}

\newcommand{\la}{\langle}
\newcommand{\ra}{\rangle}
\newcommand{\re}{\textup{Re}}

\begin{document}

\title{A Contextuality Based Quantum Key Distribution Protocol}
\author{
  J.~E.~Troupe\\Center for Quantum Research, \\Applied Research Laboratories, \\The University of Texas at Austin, \\Austin, TX 78713 \\Institute for Quantum Studies, \\Chapman University, \\Orange, CA 92866
  \texttt{jtroupe@arlut.utexas.edu}
  \and
  J.~M.~Farinholt\\Strategic and Computing Systems Department, \\Naval Surface Warfare Ctr, Dahlgren Division, \\Dahlgren, VA 22448\\
  \texttt{jacob.farinholt@navy.mil}
}
\maketitle
\begin{abstract}
In this article we present a new prepare and measure quantum key distribution protocol that uses an experimentally accessible measure of single qubit contextuality to warranty the security of the quantum channel.  The definition of contextuality used is that due to Spekkens in which any underlying hidden variable model of the physical system is noncontextual if the probability distribution of the model's hidden variables is independent of the preparation and measurement context.  Under this more general definition of noncontextuality the measurement outcomes of a single qubit can be shown to require a contextual model in order to reproduce the results of some non-projective, positive operator valued measurements (POVMs).  The proposed protocol utilizes a particular set of POVMs to exhibit the degree of contextuality of the qubits exiting the QKD system's quantum channel.  The contextuality measure is defined in terms of the weak values of a set of projective measurements that are implemented as POVMs using weak measurements.  The resulting QKD protocol should have a key rate that is equal to or greater than BB84 since none of the sifted key is used to test for the presence of an eavesdropper. Additionally, the new protocol is shown to be immune to detector based attacks. 
\end{abstract}




\section{Introduction}

The intuition that quantum mechanics provides a unique resource for information security dates back to at least the early 1970's with Wiesner's ``quantum money'' \cite{Wies83}.  This intuition was fulfilled to a large degree with the invention of quantum key distribution in 1984 \cite{BB84}, and the subsequent development of practical QKD systems.  However, there is a sense in which QKD, as currently implemented, is still unsatisfying.  This is due to the fact that if we view these systems as black boxes and we are only given the input and output data, we really have no way of telling if the information channel between Alice and Bob was truly a quantum channel, or only a noisy classical one.  Of course QKD systems do provide some level of security if we are allowed accurate information about the details of the system components.  Essentially, the security rests on the validity of this information about the QKD system implementation.  As an example, in BB84 the security relies on the assumption that Bob's detectors are and will remain equally sensitive to photons prepared in either basis.  If this assumption is violated, for example by an eavesdropper selectively controlling Bob's detector sensitivity, then the security of the protocol is broken \cite{Lyder10}.  Note that given only the input and output statistics of the system, there is no difference between the presence and absence of Eve in this example.  This is due to the fact that the input/output statistics of projective measurements on single qubits can be modeled entirely with classical random variables.  

For this reason people have begun to explore the possibility of creating QKD protocols that exploit features of quantum mechanics such that the measurement statistics are incompatible with any classical (i.e. local, realistic) model.  These are commonly known as ``Device Independent QKD'' (DI-QKD) protocols \cite{E91, GisPirSan10}.  The basis for security of these protocols is that there exists a clear experimental procedure such that the behavior of a quantum system can be distinguished from a classical one.  The ability to make such distinctions are underpinned logically by certain ``no-go'' theorems in quantum foundations that demonstrate the impossibility of certain classes of ontological models (known colloquially as ``hidden variable'' models) to reproduce the measurement outcome statistics of quantum mechanics.  The most important no-go theorem is Bell's theorem \cite{Bell04} that rules out local, realistic models.  This theorem was operationalized with the CHSH inequality \cite{CHSH69} so that two spatially separated systems can be tested for ``classicality'', i.e. consistency with a local, realistic model.  So in principle there exists an operational procedure to distinguish between a probabilistic classical system and two entangled qubits.  

There are two very significant hurdles that block the use of such no-go theorems as the basis of a secure QKD system: (1) it requires the ability to distribute sufficient numbers of entangled qubits between Alice and Bob, and (2) there is still the need to have sufficient component performance to eliminate loopholes (e.g. detection efficiency) in the CHSH-Bell tests being performed \cite{Gerh11}.  The first hurdle blocks us from using single photon protocols such as BB84, while the second calls into question the term ``device independent'' for these protocols.  Very similar hurdles exist for other new QKD protocols such as the ``Measurement Device Independent'' (MDI-QKD) protocols e.g. those of Lo \cite{LoCurQi14} and Pan \cite{Tang14}.  For example, though entanglement is not used explicitly, two-photon interference is utilized as a signature of nonclassicality, thus expanding the state space beyond single qubits.  Doing so greatly increases the difficulty of successfully distributing keys, and furthermore it is unclear what the experimental/implementation requirements are for definitively ruling out classical models of the operational statistics of such systems.  

All of this begs the question: Can we identify an operational method of distinguishing between the transmission of \emph{single} qubits through a quantum channel and classical random bits via a classical channel?  Furthermore, if we can identify such a method, how can it be implemented to place as few requirements as possible on the performance and nature of the components required?  
In this article we will present a new QKD protocol that utilizes recent results from quantum foundations to measure the ``classicality'' of a noisy quantum channel, and ensure the security of the distributed key using very few assumptions about the measuring devices used in the implementation.  In this sense this protocol is close to the ideal of an MDI-QKD protocol. In contrast to current realizations of MDI-QKD, however, this protocol should have a key rate that compares favorably to that of the original BB84 protocol.

\section{Introducing a Weak Measurement Based Protocol}

The weak value of a quantum mechanical observable was introduced by Aharonov \cite{AAD85, AV90} over two decades ago.  The weak value is experimentally obtained from the result of measurements performed upon a pre-selected and post-selected (PPS) ensemble of quantum systems when the interaction between the measurement apparatus and each system is sufficiently weak.  Unlike the standard strong measurement of a quantum mechanical observable which disturbs the measured system and ``collapses'' its state into an eigenstate of the observable, a weak measurement does not appreciably disturb the quantum system, and yields the weak value as the measured value of the observable.  This is possible because very little information about the observable is extracted in a single weak measurement.  Experimentally determining the weak value requires performing weak measurements on each member of a large ensemble of identical PPS systems and averaging the resulting values. 

Let us imagine that the measuring system, referred to as the measurement device (MD), has an initial quantum state that is a real-valued Gaussian with unit uncertainty.  Then in the limit of very weak coupling between the measuring device and the system, the weak measurement of the observable $\widehat{A}$ results in a shift in the measuring device's wavefunction so that
\begin{equation}\label{Eq:WeakValueShift}
\varphi(x) \rightarrow \varphi(x-g\re[A_w] ),
\end{equation}
where $g$ is the interaction coupling strength and $A_w$ is the weak value of the observable being measured.  

Note that the weak value of a Hermitian observable can in general be complex valued, and even when it is real valued, the weak value can lie outside its eigenspectrum.  It was recently shown that when the weak value of a projector is real and negative (i.e. lies outside the projector's eigenspectrum) then the system being weakly measured demonstrates contextuality \cite{Pus14}, and thus it cannot be completely modeled by a noncontextual ontological model \cite{Spek05}.  In particular, this implies that the system being measured cannot be modeled by a completely classical system. Note that the definition of contextuality we are using is that due to Spekkens and is a generalization of the original term contextuality introduced by Kochen and Specker \cite{KochSpek67}.  

The basis for the security of the proposed QKD protocol is that if the quantum channel demonstrates a sufficient degree of contextuality, then it is secure.  In order to do this we will define a measure of contextuality for the quantum channel using the weak values of a collection of projection operators on the Bloch sphere.  Before we define the contextuality measure, let us specify the basic protocol. We refer the reader to Table \ref{Tab:Protocol} for an outline of the protocol.

\begin{table}
\centering
\begin{tabular}{|c|l|}\hline
\  & \textbf{THE WEAK VALUE BASED QKD PROTOCOL} \\\hline\hline
\ & For a block of size $N$, Alice prepares each photon randomly in the basis $X$ or $Z$ with\\
1 & random bit values and sends each photon to Bob through the quantum channel.\\\hline
\ & For each photon, Bob weakly measures one of the four $\widehat{H}$ projectors (chosen randomly)\\
2 & and records weak measurement results in a time-ordered list.\\\hline
\ & For each photon, Bob randomly chooses the post-selection basis $X$ or $Z$, performs the \\
3 & final projective measurement, and records the results in a time-ordered list.\\\hline
\ & \ \\
4 & Alice announces her choice of preparation basis for each photon.\\\hline
\ & Bob announces which of the detected photons were post-selected in a basis \emph{different}\\
5 & from the one in which they were prepared.\\\hline
\ & \ \\
6 & Alice announces the raw key bit values for each of the photons in these cross-basis cases.\\\hline
\ & Bob separates his weak measurement results associated with these photons into the four\\
7 & different PPS ensembles and calculates the average weak measurement results of each.\\\hline
\ & Bob uses the resulting weak measurement averages for the PPS ensembles to estimate the \\
8 & weak measurement coupling parameter and the weak values of the four $\widehat{H}$ projectors.\\\hline
\ & If the weak values are sufficiently anomalous, Bob declares the channel secure. Otherwise,\\
9 & Alice and Bob abort the protocol.\\\hline
\ & Alice and Bob perform error reconciliation and privacy amplification to generate a \\
10 & secure key.\\\hline
\end{tabular}
\caption{An outline of the proposed weak value based QKD protocol.}
\label{Tab:Protocol}
\end{table}

We imagine that Alice begins with a uniformly random block of raw key of length $2N$.  For each bit of the raw key Alice will prepare the state of a single qubit and send it to Bob over the quantum channel.  The protocol will use the same four quantum states as BB84: $|0\ra$, $|1\ra$, $|+\ra$, $|-\ra$,  which represent the positive and negative eigenstates of the $\widehat{Z}$ and $\widehat{X}$ Pauli operators respectively.  These four states are used to encode the raw key bit value so that ``0'' is represented by $|0\ra$ or $|+\ra$ and ``1'' is represented by $|1\ra$ or $|-\ra$.  Alice will uniformly randomly choose one of the four states, prepare the qubit in that state and send it to Bob via the quantum channel.  Bob will perform weak measurements of one of the four following projectors on each of the qubits chosen uniformly at random
\begin{equation}
\widehat{H}^\pm \equiv \frac{1}{2}\left[ \widehat{I} + \frac{1}{\sqrt{2}}(\widehat{X} \pm \widehat{Z})\right],
\end{equation}
and the two complementary projectors 
\begin{equation}
\widehat{H}^\pm_\bot \equiv \frac{1}{2}\left[ \widehat{I} - \frac{1}{\sqrt{2}} (\widehat{X} \pm \widehat{Z})\right].
\end{equation}
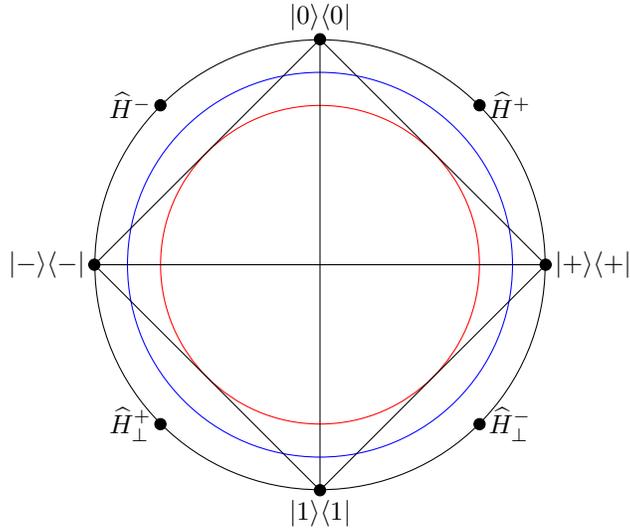
\begin{figure}%
\centering 

\begin{tikzpicture}[scale=3]
\coordinate [] (origin) at (0,0);
\coordinate [label=right:$|+\ra\la+|$] (+) at (1,0);
\coordinate [label=left:$|-\ra\la -|$] (-) at (-1,0);
\coordinate [label=right:$\widehat{H}^+$] (H+) at (45:1cm);
\coordinate [label=above:$|0\ra\la 0|$] (0) at (0,1);
\coordinate [label=below:$|1\ra\la 1|$] (1) at (0,-1);
\coordinate [label=left:$\widehat{H}^-$] (H-) at (135:1cm);
\coordinate [label=left:$\widehat{H}^+_\bot$] (H+bot) at (-135:1cm);
\coordinate [label=right:$\widehat{H}^-_\bot$] (H-bot) at (-45:1cm);

\coordinate (noise) at ($ (0)!0.5!(+) $){};
\coordinate (securebound) at ($ (noise)!0.5!(H+) $){};


\node (+node) at (+) [circle, draw, fill, inner sep=1.5pt]{};
\node (H+node) at (H+) [circle, draw, fill, inner sep=1.5pt]{};
\node (-node) at (-) [circle, draw, fill, inner sep=1.5pt]{};
\node (0node) at (0) [circle, draw, fill, inner sep=1.5pt]{};
\node (1node) at (1) [circle, draw, fill, inner sep=1.5pt]{};
\node (H-node) at (H-) [circle, draw, fill, inner sep=1.5pt]{};
\node (H+botnode) at (H+bot) [circle, draw, fill, inner sep=1.5pt]{};
\node (H-botnode) at (H-bot) [circle, draw, fill, inner sep=1.5pt]{};

\node (Circ) [draw,circle through=(H+)] at (origin) {};
\node (CircNoise) [draw,circle through=(noise), color=red] at (origin) {};
\node (CircSecure) [draw,circle through=(securebound), color=blue] at (origin) {};

\draw (0)--(+);
\draw (0)--(-);
\draw (1)--(+);
\draw (1)--(-);
\draw (0)--(1);
\draw (-)--(+);

\end{tikzpicture}%
\caption{A view of the relation between the weak measurement projectors and the projectors onto the four states used in BB84 in the space of density operators. In this space, states and projectors onto the states can be viewed as one and the same. Consequently, depolarizing noise in a QKD channel can be viewed as a uniform contraction of the outer circle towards the origin.  If the noise is large enough, then the channel can be modeled by an entirely classical model.  The smaller, red circle denotes this situation.  The blue circle denotes the noise threshold for obtaining a secure key as defined by the contextuality measure introduced in section 5. (color online)}\label{Fig:CirclePlot}%
\end{figure}
Figure \ref{Fig:CirclePlot} shows, among other things, the relation between the weak measurement projectors and the projectors onto the four states used in BB84. It can be shown \cite{FGT15} that these projectors are the optimal choice weak measurements, in the sense that they provide the most anomalous real weak values for the given selection of PPS ensembles. After Bob weakly measures the chosen projector, he randomly chooses the measurement basis for his post-selection and performs a strong measurement of the qubit.  Bob records the weak measurement results of the projector and the final state of the qubit resulting from the strong measurement.  This procedure is repeated for all of the bits in the raw key.  As will become clear below, it is important to ensure that the weak measurement coupling strengths, while not assumed to be fixed and known, \emph{a priori} are identical for all four of the projectors.  

Next, over an insecure (but authenticated) classical channel, Alice reveals the basis in which she prepared each of the qubits.  Bob then announces which of the qubits his choice of basis for the strong measurements matched Alice's basis choice.  For the qubits for which Alice's and Bob's basis choice did not match, Alice announces the raw key bit values she used to prepare the initial qubit states.  Bob then uses this information to separate his weak measurement results into distinct PPS ensembles.  Bob uses the weak measurement results from each PPS to estimate the associated weak values.  In the ideal case with no noise or eavesdropper in the channel, the weak values associated with half of the PPS ensembles are anomalous.  If the weak values are sufficiently anomalous as defined by a particular contextuality measure, then Bob declares that the quantum channel is secure and key reconciliation and privacy amplification are used to generate a secure shared key.  In the next section we will present the weak values for the various PPS in the presence of channel and detector noise. In section 5 we will define a contextuality measure and demonstrate its relation to the channel security threshold.

\section{Weak Values for Protocol PPS Ensembles}

Later in this section we will take the effect of detector dark counts into account, however for now let us assume they are zero.  Given that the quantum channel from Alice to the location of Bob's weak measurement interaction has a bit error probability of $p_a$ and the bit error probability from the weak measurement to Bob's post-selection is $p_b$, the weak values of the projectors for the PPS ensembles are then given below, calculated using the formalism of \cite{DJ12} for generalized weak values where the pre-selected and post-selected states need not be pure.  Note that the usual time symmetry of the weak values for each PPS is broken for some of the observables due to the presence of noise.  It is interesting to note that this broken symmetry only occurs for the weak values that are not anomalous for a given PPS.  

\textbf{PPS (1a)} Initial state $|0\ra$ and final state $|+\ra$:
\begin{align} \label{Eq:1aHw+}
H^{+}_w(p_a, p_b) &= \frac{1}{2} + \frac{1-(p_a+p_b )}{\sqrt{2}}\\
H^{-}_w(p_a, p_b) &= \frac{1}{2} + \frac{(p_a - p_b )}{\sqrt{2}}.
\end{align}

\textbf{PPS (1b)} Initial state $|+\ra$ and final state $|0\ra$:
\begin{align} 
H^+_w(p_a, p_b) &= \frac{1}{2} + \frac{1-(p_a+p_b )}{\sqrt{2}}\\
H^-_w(p_a, p_b) &= \frac{1}{2} - \frac{(p_a - p_b )}{\sqrt{2}}.
\end{align}

\textbf{PPS (2a)} Initial state $|1\ra$ and final state $|+\ra$:
\begin{align} 
H^+ _w(p_a, p_b) &= \frac{1}{2} + \frac{(p_a - p_b )}{\sqrt{2}}\\
H^- _w(p_a, p_b) &= \frac{1}{2} + \frac{1-(p_a+p_b )}{\sqrt{2}}.
\end{align}

\textbf{PPS (2b)} Initial state $|+\ra$ and final state $|1\ra$:
\begin{align} 
H^+ _w(p_a, p_b) &= \frac{1}{2} - \frac{(p_a - p_b )}{\sqrt{2}}\\
H^- _w(p_a, p_b) &= \frac{1}{2} + \frac{1-(p_a+p_b )}{\sqrt{2}}.
\end{align}

\textbf{PPS (3a)} Initial state $|0\ra$ and final state $|-\ra$:
\begin{align} 
H^+_w(p_a, p_b) &= \frac{1}{2} - \frac{(p_a - p_b )}{\sqrt{2}}\\
H^- _w(p_a, p_b) &= \frac{1}{2} - \frac{1-(p_a+p_b )}{\sqrt{2}}.
\end{align}

\textbf{PPS (3b)} Initial state $|-\ra$ and final state $|0\ra$:
\begin{align} 
H^+ _w(p_a, p_b) &= \frac{1}{2} + \frac{(p_a - p_b )}{\sqrt{2}}\\
H^-_w(p_a, p_b) &= \frac{1}{2} - \frac{1-(p_a+p_b )}{\sqrt{2}}.
\end{align}

\textbf{PPS (4a)} Initial state $|1\ra$ and final state $|-\ra$:
\begin{align}
H^+_w(p_a, p_b) &= \frac{1}{2} - \frac{1-(p_a+p_b )}{\sqrt{2}}\\
H^-_w(p_a, p_b) &= \frac{1}{2} - \frac{(p_a - p_b )}{\sqrt{2}}.
\end{align}

\textbf{PPS (4b)} Initial state $|-\ra$ and final state $|1\ra$:
\begin{align}
H^+_w(p_a, p_b) &= \frac{1}{2} - \frac{1-(p_a+p_b )}{\sqrt{2}}\\
H^-_w(p_a, p_b) &= \frac{1}{2} + \frac{(p_a - p_b )}{\sqrt{2}}.
\end{align}

Note that each of the four PPS ensembles labeled (b) above have the same pair of PPS states as those labeled (a) but with the time order reversed.  We have assumed that the quantum channel bit error rates are independent of the encoding basis (i.e. act as depolarizing noise).  If this is not the case, then the channel and detector error probabilities in each of the PPS cases should be additionally labeled by the encoding and measurement bases respectively.  

Detector dark counts introduce an additional parameter that is not captured above. In particular, there is the possibility that the photon Alice sends is lost prior to weak measurement, but because of detector dark counts, Bob still detects a signal. In this case, because no signal was weakly measured, the weak measurement result will have no correlation at all to the signal. This has the effect of attenuating the overall average weak value. In order to characterize this attenuation, we define a \emph{dark count attenuation parameter}, $d$, which is defined as the probability that the weak measurement result is totally uncorrelated with the signal. Thus, $d$ is given by the ratio
\begin{equation}
d: = \frac{(\text{probability that the signal was not weakly measured})}{(\text{probability of obtaining a signal})}.
\end{equation}

We note that dark counts can affect the weak measurement results in two ways depending on whether or not the photon arrived at the weak measurement location before being lost from the channel.  For the photons that make it all the way to the weak measurement, the effect of the dark count is a bit flip error on the post-selected state half of the time.  This will already be reflected in the value of $p_b$ in the weak values given above. Note that this case will occur very rarely since the probability of the photon being lost in the channel between the weak measurement and Bob's detectors is extremely small.

Another consequence of dark counts is the possibility of both of Bob's detectors clicking simultaneously. Because we are assuming that Alice is transmitting single photons, if both of Bob's detectors click simultaneously, then this either corresponds to a dark count in both detectors or to a dark count in one detector and the photon Alice sent in the other. In either case, for security reasons, Bob will uniformly at random pick one of the detectors as the signal, and ignore the other. In the former case, no signal was weakly measured, so regardless of the choice Bob makes, this will contribute an increase to the dark count attenuation parameter $d$. In the latter case, a signal was weakly measured, so it will not contribute to the parameter $d$. Bob choosing the detector that had the dark count in the latter case is effectively equivalent to the case in which a dark count occurs and the photon is lost between weak measurement and post-selection. Consequently, this will have the same effect on the weak value statistics as channel noise occurring between the weak measurement and post-selection, so it will be absorbed into the value $p_b$ above.

Now, in order to calculate the parameter $d$, let $r_{d1}$ and $r_{d2}$ denote the dark count rates of detector 1 and detector 2, respectively. If Alice had not sent a signal at all, then the probability of detecting a dark count at detector 1 in a given signal detection window is given by
\begin{equation}
p_{d1} = r_{d1}t,
\end{equation}
where $t$ corresponds to the duration of the signal detection window. We define $p_{d2}$ similarly. Consequently, if Alice had sent no signal at all, then the probability of observing a dark count signal is given by
\begin{equation}
p_{dark} = p_{d1} + p_{d2} - p_{d1}p_{d2}.
\end{equation}

The probability that a photon that Alice sends arrives at one of Bob's detectors is given by
\begin{equation} 
p_{photon} = {\eta}10^{-({\kappa}l+c)/10},
\end{equation}
where $\kappa$ denotes the channel loss rate over a distance $l$, with bulk optical loss of the system components given by $c$, and average detector efficiency given by $\eta$. We may then calculate the probability of a signal detection event to be
\begin{equation}
p_{signal} = p_{photon} + p_{dark} - (p_{photon}p_{dark}).
\end{equation}
It follows that, under the approximation that no losses occur between weak measurement and post-selection, the dark count attenuation parameter $d$ is given by
\begin{equation}
d = \frac{p_{dark}}{p_{signal}}.
\end{equation}
It follows that the observed weak value including dark count attenuation for an observable $\widehat{H}$ is 
\begin{equation}\label{Eq:Hw_DarkCounts}
H_w(p_a,p_b,d) = (1-d) H_w(p_a,p_b).
\end{equation}

For all of the PPS ensembles, the weak values of the corresponding complementary projectors are given by
\begin{equation}
(H^\pm_\bot)_w(p_a, p_b, d) = 1-H^\pm_w(p_a, p_b, d).
\end{equation}
In order to estimate the weak values for each PPS, Bob calculates the mean value of his weak measurements for each of the projectors.  The relation between the weak measurement mean values and the weak values for each PPS is 
\begin{equation}
\mu^\pm = gH^{\pm}_w(p_a,p_b,d),
\end{equation}
where, as in Eq. \eqref{Eq:WeakValueShift}, $g$ is the coupling constant. For each PPS, using the weak values $H^\pm_w$ and $(H^\pm_\bot)_w$ and their relations to the channel and detector noise given above, we can estimate the total channel noise given by $p_{channel} = p_a + p_b$ using the means of the weak measurements. When $P_{channel} = d = 0$, then for each PPS, exactly one of the $H^\pm_w$ is anomalous. We can use the mean value of the weak measurements associated with these projectors and their orthogonal complements to estimate $p_{channel}$.  Also, note that if we ensure that for each projector pair, $\widehat{H}^\pm$ and $\widehat{H}_\bot^\pm$, both are weakly measured with equal coupling strengths, then it follows that the coupling strengths are given by 
\begin{equation}
 g_\pm = \frac{\mu^\pm + \mu_\bot^\pm}{1-d}
\end{equation}
for all PPS ensembles. In fact, this relation holds for \emph{all} of the weak measurements including those for which Alice and Bob measured in the same basis.  This means that Bob can use all of his weak measurement data, in conjunction with an accurate estimate of the dark count attenuation parameter $d$, to estimate the coupling strength used for the weak measurements in the protocol independent of the intrinsic (i.e. non dark count induced) noise in the quantum channel.  Therefore, the estimate of this total channel error probability, $p_{channel}=p_a + p_b$, can be obtained directly from the mean values of the conditional weak measurement results. \\

\textbf{PPS (1):}
\begin{align}
 p_{channel} &= 1 - \frac{1}{\sqrt{2}}\left(\frac{\mu^+ - \mu^+_\bot}{\mu^+ + \mu^+_\bot}\right)
\end{align}

\textbf{PPS (2):}
\begin{align}
 p_{channel} &= 1 - \frac{1}{\sqrt{2}}\left(\frac{\mu^- - \mu_\bot^-}{\mu^- + \mu^-_\bot}\right)
\end{align}

\textbf{PPS (3):}
\begin{align}
 p_{channel} &= 1 - \frac{1}{\sqrt{2}}\left(\frac{\mu_\bot^- - \mu^-}{\mu^- + \mu^-_\bot}\right)
\end{align}

\textbf{PPS (4):} 
\begin{align}
 p_{channel} &= 1 - \frac{1}{\sqrt{2}}\left(\frac{\mu^+_\bot - \mu^+}{\mu^+ + \mu^+_\bot}\right)
\end{align}
Note that the total QBER is given by $\text{QBER} = p_{channel} + d/2$ since the errors due to dark counts when the photons do not arrive at Bob's weak measurement are not included in $p_{channel}$.  


\section{Effect of Weak Measurement Disturbance on Key Rate}

One may worry that the necessary disturbance of the weak measurements in the protocol will adversely affect the ability to generate a secure key.  After all, increases in the intrinsic error rate in the channel will decrease the rate at which a secure key can be distilled.  In this section we will estimate the QBER from state collapse due to the non-zero weak measurement interaction strength.  

Since we are focusing only on the contribution to the QBER due to the weak measurement back-action, we assume that the state just before Bob's weak measurement is pure.  Let that state be the initial qubit state $|\psi\ra = \alpha|H_\bot\ra + \beta|H\ra$ written in the $H$ basis.  Then, just after the interaction implementing the weak measurement of $|H\ra\la H|$, using a Gaussian MD wavefunction with variance $\sigma^2$ for the weak measurement pointer state, we have the following for the MD and qubit combined system,
\begin{align}
 \rho &= |\alpha|^2|H_\bot \ra\la H_\bot |\int \! e^{-x^2/4 \sigma^2} \int \! e^{-y^2/4 \sigma^2}|x\ra\la y| \mathrm{d}x \mathrm{d}y \nonumber
+  |\beta|^2|H \ra\la H |\int \! e^{-(x-g)^2/4 \sigma^2} \int \! e^{-(y-g)^2/4 \sigma^2}|x\ra\la y| \mathrm{d}x \mathrm{d}y \\ &+  \alpha\beta^*|H_\bot \ra\la H |\int \! e^{-x^2/4 \sigma^2} \int \! e^{-(y-g)^2/4 \sigma^2}|x\ra\la y| \mathrm{d}x \mathrm{d}y + \alpha^*\beta|H \ra\la H_\bot |\int \! e^{-(x-g)^2/4 \sigma^2} \int \! e^{-y^2/4 \sigma^2}|x\ra\la y| \mathrm{d}x \mathrm{d}y.
\end{align}
Tracing out the MD gives the density operator for the qubit after the weak measurement interaction,
\begin{align}
 \rho_{qubit} &=\mathrm{Tr}_{MD}(\rho)=\int \! \la x|\rho|x\ra \mathrm{d}x \nonumber  \\ 
&= |\alpha|^2|H_\bot \ra\la H_\bot| + |\beta|^2|H \ra\la H | +   e^{-g^2/8 \sigma^2} \left( \alpha\beta^*|H_\bot \ra\la H | + \alpha^*\beta|H \ra\la H_\bot | \right) \\ 
&=  e^{-g^2/8 \sigma^2}|\psi\ra\la \psi| + \left(1- e^{-g^2/8 \sigma^2}\right)\left( |\alpha|^2|H_\bot \ra\la H_\bot| + |\beta|^2|H \ra\la H | \right).
\end{align}
The last expression shows that the qubit density operator is a mixture of the undisturbed original state and collapses onto each of the two $\widehat{H}$ basis states. From this it can be easily shown that the probability for the weak measurement to collapse the original qubit state into its orthogonal state is 
\begin{equation}
p_{wm} = \frac{1}{4} \left[ 1 - \exp \left( - \frac{g^2}{8 \sigma^2} \right) \right]
\end{equation}
for all four initial qubit states and for the weak measurement of all four $H$ projectors.  This means that the weak measurement disturbance acts as a depolarizing channel with an error probability given by $p_{wm}$.  This will contribute to the observed channel error probability $p_{channel}$.  Also, while we assumed that the weak measurement pointer was a real valued Gaussian wavefunction, very similar results hold for any MD wavefunctions that are real valued and symmetric, i.e. $|\varphi(x)| = |\varphi(-x)|$, as shown by Silva, \emph{et al}. \cite{Silva15}.  

For a practical quantum optical weak measurement, the ratio of the coupling strength and MD pointer position uncertainty can be less than $0.10$.  For such a weak measurement strength, $p_{wm} < 0.0004$.  Thus, even for practical, non-zero strength weak measurements, the induced error rate from the weak measurements is less than $0.1\%$.  Note that there is a trade-off between the weak measurement disturbance induced error rate and the required raw key block size.  This is because the block size required for a fixed weak value estimation precision grows as $(\sigma/g)^2$ when the coupling strength is reduced.  This indicates that the new protocol will require raw key block sizes approximately $(\sigma/g)^2$ times larger than standard BB84 to yield equal precision of the channel QBER estimate.  Of course, the precision needed for a given level of security will depend on how close the contextuality measure (and thus the QBER) is to the secure limit.  For a quantum channel near the secure boundary, a high degree of precision is needed to ascertain if the channel is secure.  It is also important to note that (1) the new protocol provides approximately twice the key production rate of BB84 since none of the sifted key is used for QBER estimation, and (2) unlike standard BB84, the new protocol's QBER estimate is not susceptible to detector based hacking, as will be evident later. 


\section{A Contextuality Measure}

We could simply use the estimates of the QBER taken from the observed weak values to bound Eve's information about the sifted key.  However, the key question is whether we can trust that these estimates are accurate.  Is it possible that Eve biased the statistics using some flaw in the hardware?  As an example of what could happen, the QBER estimators used in BB84 can be biased by detector blinding attacks to hide Eve's presence, as in, e.g. \cite{Lyd10}.  In order to ensure that Eve could not succeed with similar strategies, we would like some measure of the nonclassicality or `quantumness' of the channel.  We need some experimental signature that rules out some degree of classicality.  

In 2005 Spekkens \cite{Spek05} introduced a generalized characterization of Kochen-Specker (K-S) noncontextuality \cite{KochSpek67} of ontological (or less precisely, hidden variable) models of quantum theory that applies to measurements that are non-projective.  Unlike K-S noncontextuality, Spekkens considers indeterministic models where the underlying hidden variable only specifies a probability distribution for the outcomes of measurements.  Two different procedures $M$, $M^\p$ for implementing the measurement of an observable are operationally equivalent if the probability distribution for the outcomes $k$ of the measurement are identical for all identical system preparations $P$, that is,

\begin{equation}
p(k | P,M)=p(k | P,M' ) \text{ for all } P.
\end{equation}

Different (but operationally equivalent) measurement procedures are called \emph{measurement contexts}.  An ontological model contains underlying ontological variables $\lambda$ specifying an indicator function $\xi _{M,k}(\lambda)$ that gives the probability for the outcomes $k$ of a given measurement procedure $M$.  Therefore, by Spekkens' definition, an ontological model is measurement noncontextual if and only if 
\begin{equation}
\xi_{M,k} (\lambda)=\xi_{e(M),k} (\lambda),
\end{equation}
where $e(M)$ is the equivalence class of measurement procedures as defined above.  In other words, a model is noncontextual if the indicator function connecting the ontological variable and the measurement outcomes is unchanged by the choice of measurement contexts.  This definition is a generalization of the more traditional Kochen-Specker definition recovered as the special case where the measurement procedures specify sharp, projective measurements and the indicator functions are all deterministic such that the probability distributions $\xi_{M,k} (\lambda)$  are idempotent (i.e. only have values ${0,1}$).  It is important to note that Spekkens' definition allows unsharp measurement procedures.  The significance of this is that it opens the possibility of empirical tests that could rule out noncontextual models of observed outcomes on realistic experiments in a way analogous to CHSH tests of Bell locality.  Of particular relevance to QKD, it was shown that non-projective measurements (POVMs) on single qubits require contextual models.  This is in contrast with the common belief that single qubits can be completely modeled by probabilistic classical (and hence noncontextual) systems.  

We propose to use the normalized size of the anomalous weak values in the PPS ensembles above as a contextuality measure.  Pusey \cite{Pus14} recently showed that for pre- and post-selected weak measurements of projectors on the Bloch sphere yielding real weak values that exceed the bounds of $[0,1]$, a qubit cannot be modeled by a noncontextual ontological model as defined by Spekkens.  In particular, this means that there is no classical model of the qubit's state that accurately reproduces all of the weak measurement outcomes indicative of the anomalous weak values of the proposed protocol under the assumption that the encoded states are single photons.  The assumption of single qubits is needed for the same reason that it is needed in the original BB84 protocol, namely because any redundancy in encoding the raw key reduces the security independent of any properties of the single qubit quantum channel.  For any practical implementation of the proposed protocol, some countermeasures must be taken to limit such side-channel information available to Eve.  Modification of the proposed protocol to incorporate such countermeasures will be the subject of future research.  

Intuitively, the closer a noisy qubit is to being classical, the more information Eve could in principle have, since she could have a classical model that allows her to predict the measurement outcomes. 
If all the noise is depolarizing, then the red circle in Figure \ref{Fig:CirclePlot} indicates the point at which the noise can be modeled using a purely classical model. With this ability, Eve could then use her knowledge of the quantum state and any imperfections in the QKD system to mask her presence and break the protocol's security. Consequently, the degree of noncontextuality provides a measure of how much information Eve can potentially obtain.  By bounding the degree of noncontextuality in the noisy qubits exiting the channel, we limit the capability of Eve to model the qubit's behavior and extract key information while evading detection, even when she has control of Bob's detectors.  

We will think of the contextuality of the qubits as a physical resource, and note that by traveling through a noisy quantum channel, that resource is degraded.  This is evidenced by the attenuation of the weak values as a function of the channel and detector noise.  Given an anomalous weak value for a projector $\widehat{H}$, i.e. $H_w (p_{channel}, d) \not\in [0,1]$, where again $p_{channel}$ is the total error probability of the quantum channel and $d$ is the dark count attenuation parameter, we define the measure of contextuality of a qubit to be
\begin{equation}\label{Eq:ContextualityMeasure}
\mc{C}(p_{channel},d) = \frac{|\re[H_w (p_{channel},d)]- \text{near}(\widehat{H})|}{|\re[H_w(0,0)]- \text{near}(\widehat{H})|},
\end{equation}
where $\text{near}(\widehat{H})$ is the nearest eigenvalue of the projector $\widehat{H}$ to the real part of the anomalous weak value $H_w (p_{channel}, d)$.  Essentially, this is the normalized distance between the anomalous weak value of the noisy qubit and the nearest eigenvalue bound of the observable.  Note that this measure is only defined when the weak value is outside the eigenvalue range of the projector. To complete the definition, we will set $\mc{C}=0$ whenever $\re[H_w(p_{channel}, d)] \in [0,1]$.  

When $\mc{C}=1$ there is no noise in the quantum channel and no dark counts, and the qubits will exhibit a maximum of contextuality.  In this case the channel is purely quantum, with no noise and no classicality.  Consequently, Eve has no access to the information in the channel. When the noise is large enough,  $\mc{C}=0$ and the channel can be completely classical.  In this case Eve can in principle have full access to all of the information in the channel.  If we assume that Eve attacks the channel only with individual attacks with no entanglement across channel uses, we can measure the contextuality of a general, noisy quantum channel by a convex combination of these two extrema:
\begin{equation}
\mc{C}(p_{channel},d)=(1-a) \mc{C}_{quantum} + a\mc{C}_{classical}.
\end{equation}

Essentially, we are treating the noisy quantum channel as a statistical mixture of a completely secure noise-free quantum channel and a completely insecure channel that can in principle be modeled by a noncontextual (i.e. classical) channel.  We can do this because all parties involved (including Eve) are treating each use of the channel independently, with e.g. no use of entanglement across uses of the channel. To find the maximum secure QBER for Eve's individual attacks, we set the mixture so that there are equal classical and quantum proportions, thus $a=1/2$.  At this point Eve could have access to as much information about the raw key as Alice and Bob. This yields a secure bound of $\mc{C}(p_{channel},d)>1/2$.  We can therefore certify the quantum channel secure against individual attacks if and only if the weak measurement results clearly indicate that the contextuality measure exceeds this lower bound.  

To demonstrate that this contextuality bound is consistent with known limits for quantum channel error rate, we take as an example the anomalous weak value from Eq. \eqref{Eq:1aHw+}. Then using Equations \eqref{Eq:Hw_DarkCounts} and \eqref{Eq:ContextualityMeasure}, we obtain: 
\begin{equation}\label{Eq:ContextMeasure}
\mc{C}(p_{channel},d) = \frac{ (1-d) \left(\frac{1}{2}+\frac{1 - p_{channel}}{\sqrt{2}}\right) - 1}{ \frac{1}{\sqrt{2}} - \frac{1}{2}} > \frac{1}{2}.
\end{equation}
From this we get a bound relating the channel error probability and the dark count attenuation, given by
\begin{equation}\label{Eq:PchannelDark}
p_{channel}(1-d)+d\left(1 + \frac{1}{\sqrt{2}}\right) < \frac{1}{2} -\frac{\sqrt{2}}{4}.  
\end{equation}
The same result is obtained by choosing any of the weak values which are anomalous in each of the PPS ensembles.

We note that if we set the dark count rate to zero, we get $p_{channel}< \frac{1}{2} -\frac{\sqrt{2}}{4} \approx 0.1464466$, which is the well-established QBER error threshold for BB84 under optimal individual attacks \cite{Fuchs97}. Equation \eqref{Eq:PchannelDark} also reveals that, regardless of the channel error rate $p_{channel}$, the inequality  in Equation \eqref{Eq:ContextMeasure} will always be violated when $d$ satisfies
\begin{equation}
d \geq \frac{\sqrt{2} - 1}{2(\sqrt{2}+1)} \approx 0.0857864.
\end{equation}
Consequently, the security of this protocol is not necessarily ensured by simply specifying the QBER since channel errors and dark count errors do not have an equal impact on the observed contextuality.

To indicate the feasibility of the protocol we will now sketch one particular implementation.  Let us suppose that the raw key information is encoded in the polarization of single photons with Alice randomly choosing to encoded in either the linear polarization basis ($Z$) or the diagonal basis ($X$). Furthermore, the weak measurement pointer will be the temporal wavefunction of the single photons. The weak measurement pointer states are prepared using a shutter so the single photons are emitted with a temporal envelope that is Gaussian shaped with a fixed width $\tau$ and a known mean determined by an accurate clock. Bob will perform the post-selection measurements by randomly choosing between the two polarization bases by either rotating the polarization by $\pi/4$ or not before performing single photon detection at the outputs of a polarizing beam splitter aligned with the linear basis. Bob will have a clock that is synchronized with Alice's clock.  Bob uses this clock to record the detection time of each photon.  In the new protocol Bob must also weakly measure the projectors onto the four $H$ polarization states given by 

\begin{align}
\vert H^\pm \rangle = \cos\left(\pm \frac{\pi}{8}\right)|0\rangle + \sin\left(\pm \frac{\pi}{8}\right)|1\rangle
\end{align}

\begin{align}
\vert H^{\pm}_{\perp} \rangle = \cos\left(\pm \frac{\pi}{8} + \frac{1}{2}\right)|0\rangle + \sin\left(\pm \frac{\pi}{8} + \frac{1}{2}\right)|1\rangle.
\end{align}

To implement the necessary weak measurements, Bob will have a polarizing interferometer consisting of two polarizing beam splitters oriented with the linear basis and a polarization independent optical delay in one path of the interferometer with a duration of $\delta << \tau$, so that the photon's path information is not determined by the interaction with the interferometer.  Just before the photon enters the polarizing interferometer, Bob will uniformly randomly choose to measure one of the four projectors above by rotating the incoming photon's polarization state by ${\frac{\pi}{8}, \frac{-\pi}{8}, \frac{5\pi}{8}, \frac{3\pi}{8} }$ for ${H^+, H^-, H^{+}_{\perp}, H^{-}_{\perp} }$ respectively. On exiting the interferometer, the polarization of each photon is rotated back to its original state.  Note that by implementing the weak measurement interaction in this way, the interaction strength is ensured to always be the same for all four projectors since the same delay is used.  The weak measurement pointer used here is the time degree of freedom for the photons' exiting of the source aperture.  As long as the shutter and detector timing jitter are relatively small compared to the weak measurement's optical delay, accurate estimates of delay time and the weak values can be obtained using the pre- and post-selected timing information as prescribed above.  In addition to the timing information, the detector dark count probability should be independently estimated at frequent intervals.  This could be done by having Bob randomly choose to block the photon's path to the detectors and record the rate at which the detectors fire.  In the worst (most paranoid) case, this would be done for half of the signals sent by Alice.  Note that this would then reduce the secure key rate by a factor of 1/2, making the secure rate essentially identical to the original BB84 protocol -- without, as we will see in the next section, BB84's susceptibility to detector based attacks.  


\section{Efficacy of Detector Blinding Against the Protocol}

A general method of detector-based hacking of BB84 and similar prepare-and-measure-based QKD protocols is to perform an intercept-resend attack on the quantum channel and simultaneously blind Bob's detectors so that they only respond when Bob's choice of measurement basis happens to match the random choice made by Eve during the intercept-resend measurement.  The result of this is that when Alice and Bob use a portion of the sifted key to estimate the QBER, Eve's intervening measurements will have no effect on the statistics.  A key point is that for the detector attack to work, Eve and Bob's measurement bases must match.  

Let us assume that Eve has complete control of Bob's detector, and she performs an intercept-resend attack on all of the qubits in the channel at some point prior to Bob's weak measurement. Because Eve and Bob's bases will agree, the weak values for all of the projectors must equal their expectation values.  In particular, this means that the weak measurement results for all of the PPS ensembles will equal the expectation values with respect to Bob's final measurement results (which will also match Eve's preparation states).  Because these are expectation values, they must fall strictly within the eigenvalue spectrum of the projectors. Also notice that the symmetry in the PPS ensemble weak measurement results with respect to swapping Alice's and Bob's PPS states is now broken.  The mean values of the weak measurement results associated with Bob's four strong measurement outcomes are $\mu = g_\pm(H^\pm)_w$ with the weak values given by the following: 

\noindent Bob's strong measurement result is $|0\ra$:
\begin{align}
H^+ _w &= \frac{1}{2} + \frac{1}{2\sqrt{2}}\\
H^- _w &= \frac{1}{2} - \frac{1}{2\sqrt{2}}
\end{align}

\noindent Bob's strong measurement result is $|1\ra$:
\begin{align}
H^+_w &= \frac{1}{2} - \frac{1}{2\sqrt{2}}\\
H^-_w &= \frac{1}{2} + \frac{1}{2\sqrt{2}}
\end{align}

\noindent Bob's strong measurement result is $|+\ra$:
\begin{align}
H^+_w &= \frac{1}{2} + \frac{1}{2\sqrt{2}}\\
H^-_w &= \frac{1}{2} + \frac{1}{2\sqrt{2}}
\end{align}

\noindent Bob's strong measurement result is $|-\ra$:
\begin{align}
H^+_w &= \frac{1}{2} - \frac{1}{2\sqrt{2}}\\
H^-_w &= \frac{1}{2} - \frac{1}{2\sqrt{2}}.
\end{align}

When Bob checks the mean values of the weak measurement results of the eight PPS ensembles in the protocol, he will see that they are given by the results above.  The contextuality measure from Eq. \eqref{Eq:ContextualityMeasure} of all PPS are zero, and furthermore, there is a distinctive signature given by the weak measurement results that indicate a detector attack on the protocol.  Thus, detector blinding attacks actually increase the ability of Alice and Bob to identify Eve, in contrast to other protocols.   

\section{Conclusion and Future Research}

We have presented a new single qubit prepare and measure QKD protocol that bounds a potential eavesdropper's information about the distributed key using a weak measurement derived contextuality measure.  The protocol does not use any of the sifted key to estimate the QBER of the channel or detect a potential eavesdropper, leaving all of it available for key distillation.  Additionally, the protocol is naturally immune to single-photon detector based hacking strategies. Finally, the secure key rate of a practical implementation of this protocol should compare quite favorably to standard prepare and measure protocols such as BB84.  Future work will explore extending the new protocol to perform securely with non-ideal single photon sources such as weak coherent states.

This is the first QKD protocol to base its security directly on the operational demonstration of contextuality, and opens up a new avenue of research with respect to the connection between quantum foundations tools and quantum information science.  Many interesting questions remain to be answered about contextuality measures and quantum information applications of them.  For example, when we move to higher dimensional quantum information state spaces such as qudits, what are the optimal observables to use in order to detect and quantify the degree of contextuality of a quantum system? What role can operational contextuality measures play in building better quantum error correction codes and quantum error suppression techniques?  We look forward to pursuing answers to these questions in the future.  

\section*{Acknowledgments}

J. Troupe acknowledges support from the Office of Naval Research, grant number N00014-15-1-2225 and thanks the Institute for Quantum Studies at Chapman University for their hospitality.  J. Farinholt acknowledges support from the Quantum Information Science Program, Office of Naval Research Code 31.

\bibliographystyle{unsrt}
\bibliography{bibfile3}

\begin{thebibliography}{10}

\bibitem{Wies83}
Stephen Wiesner.
\newblock Conjugate coding.
\newblock {\em SIGACT News}, 15(1):78--88, January 1983.

\bibitem{BB84}
Charles Bennett and Gilles Brassard.
\newblock Quantum cryptography: Public key distribution and coin tossing.
\newblock In {\em Proceedings of the IEEE International Conference on
  Computers, Systems, and Signal Processing}, page 175, 1984.

\bibitem{Lyder10}
Lars Lydersen, Carlos Wiechers, Christoffer Wittmann, Dominique Elser, Johannes
  Skaar, and Vadim Makarov.
\newblock Hacking commercial quantum cryptography systems by tailored bright
  illumination.
\newblock {\em Nature Photonics}, 4:686--689.

\bibitem{E91}
Artur~K. Ekert.
\newblock Quantum cryptography based on bell's theorem.
\newblock {\em Phys. Rev. Lett.}, 67:661--663, Aug 1991.

\bibitem{GisPirSan10}
Nicolas Gisin, Stefano Pironio, and Nicolas Sangouard.
\newblock Proposal for implementing device-independent quantum key distribution
  based on a heralded qubit amplifier.
\newblock {\em Phys. Rev. Lett.}, 105:070501, Aug 2010.

\bibitem{Bell04}
John~Stewart Bell.
\newblock {\em Speakable and unspeakable in quantum mechanics: Collected papers
  on quantum philosophy}.
\newblock Cambridge university press, 2004.

\bibitem{CHSH69}
John~F. Clauser, Michael~A. Horne, Abner Shimony, and Richard~A. Holt.
\newblock Proposed experiment to test local hidden-variable theories.
\newblock {\em Phys. Rev. Lett.}, 23:880--884, Oct 1969.

\bibitem{Gerh11}
Ilja Gerhardt, Qin Liu, Ant\'{i}a Lamas-Linares, Johannes Skaar, Valerio
  Scarani, Vadim Makarov, and Christian Kurtsiefer.
\newblock Experimentally faking the violation of bell's inequalities.
\newblock {\em Phys. Rev. Lett.}, 107:170404, Oct 2011.

\bibitem{LoCurQi14}
Hoi-Kwong Lo, Marcos Curty, and Bing Qi.
\newblock Measurement-device-independent quantum key distribution.
\newblock {\em Phys. Rev. Lett.}, 108:130503, Mar 2012.

\bibitem{Tang14}
Yan-Lin Tang, Hua-Lei Yin, Si-Jing Chen, Yang Liu, Wei-Jun Zhang, Xiao Jiang,
  Lu~Zhang, Jian Wang, Li-Xing You, Jian-Yu Guan, Dong-Xu Yang, Zhen Wang, Hao
  Liang, Zhen Zhang, Nan Zhou, Xiongfeng Ma, Teng-Yun Chen, Qiang Zhang, and
  Jian-Wei Pan.
\newblock Measurement-device-independent quantum key distribution over 200 km.
\newblock {\em Phys. Rev. Lett.}, 113:190501, Nov 2014.

\bibitem{AAD85}
David~Z. Albert, Yakir Aharonov, and Susan D'Amato.
\newblock Curious new statistical prediction of quantum mechanics.
\newblock {\em Phys. Rev. Lett.}, 54:5--7, Jan 1985.

\bibitem{AV90}
Yakir Aharonov and Lev Vaidman.
\newblock Properties of a quantum system during the time interval between two
  measurements.
\newblock {\em Phys. Rev. A}, 41:11--20, Jan 1990.

\bibitem{Pus14}
Matthew~F. Pusey.
\newblock Anomalous weak values are proofs of contextuality.
\newblock {\em Phys. Rev. Lett.}, 113:200401, Nov 2014.

\bibitem{Spek05}
R.~W. Spekkens.
\newblock Contextuality for preparations, transformations, and unsharp
  measurements.
\newblock {\em Phys. Rev. A}, 71:052108, May 2005.

\bibitem{KochSpek67}
Simon Kochen and E.~P. Specker.
\newblock {T}he {P}roblem of {H}idden {V}ariables in {Q}uantum {M}echanics.
\newblock {\em {J}ournal of {M}athematics and {M}echanics}, 17:59--87, 1967.

\bibitem{FGT15}
Jacob Farinholt, Alan Ghazarians, and James Troupe.
\newblock The geometry of qubit weak values.
\newblock {\em arXiv:1512.02113 [quant-ph]}.

\bibitem{DJ12}
J.~Dressel and A.~N. Jordan.
\newblock Contextual-value approach to the generalized measurement of
  observables.
\newblock {\em Phys. Rev. A}, 85:022123, Feb 2012.

\bibitem{Silva15}
Ralph Silva, Nicolas Gisin, Yelena Guryanova, and Sandu Popescu.
\newblock Multiple observers can share the nonlocality of half of an entangled
  pair by using optimal weak measurements.
\newblock {\em Phys. Rev. Lett.}, 114:250401, Jun 2015.

\bibitem{Lyd10}
Lars Lydersen, Carlos Wiechers, Christoffer Wittmann, Dominique Elser, Johannes
  Skaar, and Vadim Makarov.
\newblock Hacking commercial quantum cryptography systems by tailored bright
  illumination.
\newblock {\em Nature Photonics}, 4:686 -- 689, 2010.

\bibitem{Fuchs97}
Christopher~A. Fuchs, Nicolas Gisin, Robert~B. Griffiths, Chi-Sheng Niu, and
  Asher Peres.
\newblock Optimal eavesdropping in quantum cryptography. i. information bound
  and optimal strategy.
\newblock {\em Phys. Rev. A}, 56:1163--1172, Aug 1997.

\end{thebibliography}

\section{Appendix A}

In addition to estimating the total channel error probability $p=p_a + p_b$, the weak measurements in the new QKD protocol provide estimates for each of these error sources independently.  For completeness, here we report the relation between the mean weak measurement results conditioned on the eight PPS ensembles and the error probability due to the quantum channel from Alice's source to the weak measurement and the quantum channel from the weak measurement to Bob's detector, $p_a$ and $p_b$ respectively.  While not strictly necessary for securing the quantum channel, access to estimates of these parameters continuously while the QKD system is operational can be very useful for monitoring system performance, and gives extra situational awareness for any changes in the system's behavior.  

We remind the reader that the observed weak values for each PPS are related to the conditional weak measurement results by the relation
\begin{align} 
H^\pm_w &= \frac{\mu^\pm}{\mu^\pm + \mu_\bot^\pm}.
\end{align}
Next we will relate the channel and detector errors to the observed weak values of $\widehat{H}^\pm$ for each PPS.  

\textbf{PPS (1a)} Initial state $|0\ra$ and final state $|+\ra$:
\begin{align}
p_b &= \frac{1 + \sqrt{2} }{2} - \frac{1}{\sqrt{2}} \left(H^+_w + H^-_w \right)\\
p_a &= \frac{1}{2} - \frac{1}{\sqrt{2}} \left(H^+_w - H^-_w \right).
\end{align}

\textbf{PPS (1b)} Initial state $|+\ra$ and final state $|0\ra$:
\begin{align} 
p_b &=  \frac{1}{2} - \frac{1}{\sqrt{2}} \left(H^+_w - H^-_w \right)\\
p_a &= \frac{1 + \sqrt{2} }{2} - \frac{1}{\sqrt{2}} \left(H^+_w + H^-_w \right).
\end{align}

\textbf{PPS (2a)} Initial state $|1\ra$ and final state $|+\ra$:
\begin{align} 
p_b &= \frac{1 + \sqrt{2} }{2} - \frac{1}{\sqrt{2}} \left(H^+_w + H^-_w \right)\\
p_a &= \frac{1}{2} + \frac{1}{\sqrt{2}} \left(H^+_w - H^-_w \right).
\end{align}

\textbf{PPS (2b)} Initial state $|+\ra$ and final state $|1\ra$:
\begin{align} 
p_b &= \frac{1}{2} + \frac{1}{\sqrt{2}} \left(H^+_w - H^-_w \right)\\
p_a &= \frac{1 + \sqrt{2} }{2} - \frac{1}{\sqrt{2}} \left(H^+_w + H^-_w \right).
\end{align}

\textbf{PPS (3a)} Initial state $|0\ra$ and final state $|-\ra$:
\begin{align} 
p_b &= \frac{1 - \sqrt{2} }{2} + \frac{1}{\sqrt{2}} \left(H^+_w + H^-_w \right)\\
p_a &= \frac{1}{2} - \frac{1}{\sqrt{2}} \left(H^+_w - H^-_w \right).
\end{align}

\textbf{PPS (3b)} Initial state $|-\ra$ and final state $|0\ra$:
\begin{align} 
p_b &= \frac{1}{2} - \frac{1}{\sqrt{2}} \left(H^+_w - H^-_w \right)\\
p_a &= \frac{1 - \sqrt{2} }{2} + \frac{1}{\sqrt{2}} \left(H^+_w + H^-_w \right).
\end{align}

\textbf{PPS (4a)} Initial state $|1\ra$ and final state $|-\ra$:
\begin{align}
p_b &= \frac{1 - \sqrt{2} }{2} + \frac{1}{\sqrt{2}} \left(H^+_w + H^-_w \right)\\
p_a &= \frac{1}{2} + \frac{1}{\sqrt{2}} \left(H^+_w - H^-_w \right).
\end{align}

\textbf{PPS (4b)} Initial state $|-\ra$ and final state $|1\ra$:
\begin{align}
p_b &= \frac{1}{2} + \frac{1}{\sqrt{2}} \left(H^+_w - H^-_w \right)\\
p_a &= \frac{1 - \sqrt{2} }{2} + \frac{1}{\sqrt{2}} \left(H^+_w + H^-_w \right).
\end{align}

\end{document}